\documentstyle[12pt]{article}

\newcommand{\reseteqnum}{\setcounter{equation}{0}}

\newcommand{\be}{\begin{equation}}
\newcommand{\ee}{\end{equation}}
\newcommand{\bea}{\begin{eqnarray}}
\newcommand{\eea}{\end{eqnarray}}

\catcode`\@=11
\newbox\tempboxa
\newdimen\captionboxsubcount
\def\capsize#1{\captionboxsubcount=#1pt}
\newdimen\captionboxsub
\captionboxsub=\hsize \advance\captionboxsub by -\captionboxsubcount
\advance\captionboxsub by -\captionboxsubcount
\long\def\@makecaption#1#2{
 \setbox\@tempboxa\hbox{#1: #2}
 \ifdim \wd\@tempboxa >\captionboxsub
\rightskip=\captionboxsubcount \leftskip=\captionboxsubcount #1: #2
\else \hbox to\hsize{\hfil\box\@tempboxa\hfil}
 \fi}
\catcode`\@=12
\capsize{30}

\begin{document}

\begin{titlepage}
\begin{flushright}
hep-th/9903240 \\
\end{flushright}
\bigskip
\bigskip

\begin{center} \Large\bf 
ASYMPTOTIC SYMMETRY AND\\
THE GENERAL BLACK HOLE SOLUTION\\
IN $\bf AdS_3$ GRAVITY
\end{center}
\bigskip

\begin{center} \large
   YUHSUKE YOSHIDA
\quad and \quad
   TAKAHIRO KUBOTA
\end{center}
\bigskip

\begin{center} \large \it
         Department of Physics \\
         Graduate School of Science, Osaka University \\
         Machikaneyama 1-1, Toyonaka \\
         Osaka 560-0043, JAPAN \\
\end{center}

\begin{center} 
\Large \bf Abstract
\end{center}
\begin{quote}
We describe the Brown-Henneaux asymptotic symmetry of the general black 
holes in the Chern-Simons gauge theory of the gauge group 
$SL(2;{\bf R})_L\times SL(2;{\bf R})_R$.
We make it clear that the vector-like subgroup $SL(2; {\bf R})_{L+R}$ 
plays an essential role in describing the asymptotic symmetry
consistently.
We find a quite general black hole solution in the $AdS_3$ gravity
theory. The solution is specified by an infinite number of conserved 
quantities which constitute a family of mapping from $S^1$ 
to the gauge group. The BTZ black hole is one of the simplest case.
\end{quote}

\bigskip
\begin{flushleft}
\end{flushleft}
\end{titlepage}


\section{Introduction}\label{Intro}
\reseteqnum

There has been recently a surge of renewed interest in gravity and 
supergravity in anti-de Sitter ($AdS$) space. 
This was kindled by remarkable
observations\cite{Maldacena,GKP,Klebanov,GKT,GK,MS,GKPoly,Witten0}
that $AdS$ gravity in $(d+1)$-dimensions would describe the
$d$-dimensional conformal field theory on its boundary.
In the case of $AdS_{3}$ gravity, 
the occurrence of the two-dimensional conformal symmetry was
elucidated by Brown and Henneaux\cite{BH} more than a decade ago.
Their analysis, which is based on the work of Regge and
Teitelboim\cite{RT}, has been investigated by using the Chern-Simons
gauge theory\cite{AT,Witten} to a considerable
extent\cite{Banados,Carlip,Ezawa,OP}.
The salient feature of the Chern-Simons approach is that, although the
gauge fields do not have local excitations, the gauge degrees of
freedom on the boundary have non-trivial physical contents.

In the present paper we reanalyze the $AdS_3$ gravity.
We will work with the Chern-Simons gauge theory with the gauge group
$G=SO(2,2)=SL(2;{\bf R})_L\times SL(2;{\bf R})_R$ throughout.
We will pay a particular attention to  the properties of
diffeomorphism together with the gauge symmetry.   
To study the asymptotic symmetry of Brown and Henneaux one has to look
for the diffeomorphism that leaves the metric $g_{mn}$ unchanged.
 In the dreibein formalism, on the other hand,
the gauge transformation should also be taken into account in addition
to the diffeomorphism to derive the asymptotic symmetry.
As we will see, this is because the metric $g_{mn}$ is invariant under
the gauge transformation in the vector-like subgroup
$H=SL(2;{\bf R})_{L+R}\subset G$, while the dreibein $e_{m}^{a}$ is not. 
This means that we are necessarily led to consider a combination
of diffeomorphism ($\delta_{_D}$) and the gauge transformation
($\delta_{_H}$) in $H$.
The isometry condition is now
\begin{eqnarray}\label{imc}
(\delta_{_D}+\delta_{_H})e_m^a = 0 .
\end{eqnarray}
We will show that the asymptotic symmetry is obtained by solving
eq.~(\ref{imc}) on the boundary at infinity.
The solution of the isometry condition (\ref{imc}) at infinity
contains infinite degrees of freedom and the asymptotic symmetry is
thus infinite dimensional.
A subtle feature of our analysis in contrast to previous works is that
we consider transformations $\delta_{_D}$ and $\delta_{_H}$
simultaneously, while a full use of the subgroup $H$ and $\delta_{_H}$
eluded previous authors.

It is expected that the generators of the above asymptotic symmetry
are materialized by the Kac-Moody currents.
There has been a tacit understanding in $AdS_3$ gravity theory that
the representation space of the asymptotic symmetry bears close
connections with the black hole and vacuum solutions.
We will in fact uncover a quite general class of black hole solutions,
to which the celebrated BTZ black hole\cite{BTZ} and vacuum solutions
belong as the simplest cases.
As will be seen, the black hole solutions are characterized by
mappings from $S^1$ onto the gauge group $G$; the Kac-Moody currents.

The present paper is organized as follows.
In sect.~\ref{AdSCS}, the Chern-Simons formulation to the three
dimensional gravity is briefly summarized.
The boundary conditions for the gauge fields and the surface terms in
the action are specified.
The asymptotic symmetry on the boundary are derived by solving
eq.~(\ref{imc}) in sect.~\ref{ASAI}.
The Kac-Moody and Virasoro algebras are worked out in sect.~\ref{CACS}
and the classical part of the Virasoro central charge is derived in
agreement with the previous analyses\cite{BH}.
It is shown that the Brown-Henneaux asymptotic symmetry is associated 
not with the usual Sugawara form but with the twisted Sugawara form. 
In sect.~\ref{WZW} general black hole solutions are derived quite
generally.
Sect.~\ref{QCS} is devoted to the brief discussion of the 
quantum calculation of the Virasoro central charge.  
We will summarize our results in sect.~\ref{DIS}.

\section{$\bf AdS_3$ gravity and Chern-Simons theory}
\label{AdSCS}
\reseteqnum

In this paper we consider the three-dimensional gravity with a
negative cosmological constant.
The Einstein-Hilbert action is
\be\label{EH}
I_{EH}(g_{mn}) = \frac{1}{16\pi G_{_N}}\int
d^3x\sqrt{-g}(R+\frac{2}{\ell^2}) .
\ee
This system has the BTZ black hole solution\cite{BTZ}.
This system also has the asymptotic symmetry, which was first derived
in ref.~\cite{BH}.
One of our main purpose is to derive the asymptotic symmetry in the
BTZ black hole background using the Chern-Simons formulation with a
closer look at the role of the subgroup $H$.

In this section we review the three-dimensional gravity described by
eq.~(\ref{EH}).
Let us rewrite the action (\ref{EH}) in the dreibein formulation
\be\label{EH2}
I_{EH}(e,\omega) = -\frac{1}{8\pi G_{_N}}\int
\left(e_aR^a+\frac{1}{6\ell^2}\epsilon_{abc}e^ae^be^c\right)
+
\frac{1}{16\pi G_{_N}}\oint e_a\omega^a .
\ee
To avoid confusion we fix our notation now.
We define the three-dimensional coordinate as $(t,\phi,\rho)$,
the complete antisymmetric tensors as
$\epsilon^{t\phi\rho} = \epsilon^{012} = +1$ and
the flat metric and the local Lorentz frame metric as
$\eta_{mn} = \eta_{ab} = {\rm diag}(-,+,+)$.
We consider the system in the cylinder ${\bf R}\times\Sigma$ where
${\bf R}$ is parameterized by the time $t$, the disc $\Sigma$ is
parameterized by the angle and radial variables $(\phi,\rho)$ and,
especially, the boundary ${\bf R}\times\partial\Sigma$ is parameterized
by $(t,\phi)$.

The first term of the action (\ref{EH2}) is rewritten
as the Chern-Simons action with the gauge group
$G=SO(2,2) = SL(2;{\bf R})_L \times SL(2;{\bf R})_R$.\cite{Witten}
Both $SL(2;{\bf R})_L$ and $SL(2;{\bf R})_R$ generators are
represented by
\be\label{J_a}
J_0 = \frac{1}{2}\pmatrix{0 & -1 \cr 1 & 0\cr},\quad
J_1 = \frac{1}{2}\pmatrix{0 & 1 \cr 1 & 0\cr},\quad
J_2 = \frac{1}{2}\pmatrix{1 & 0 \cr 0 & -1\cr}.
\ee
Then, we have $[J_a,J_b] = \epsilon_{ab}{}^cJ_c$,
${\rm tr}J_aJ_b = \eta_{ab}/2$.
The combinations $A^a_m\equiv\omega^a_m+e^a_m/\ell$ and
$B^a_m\equiv\omega^a_m-e^a_m/\ell$ become
the $SL(2;{\bf R})_L$ and $SL(2;{\bf R})_R$ gauge fields,
respectively.
The Chern-Simons action is
\be\label{csa}
I_{CS}(A) = \frac{k}{4\pi}{\rm tr}\int
\left(AdA+\frac{2}{3}A^3\right) .
\ee
Then, the first term of the action (\ref{EH2}) becomes
$I_{CS}(B) - I_{CS}(A)$ with $k=\ell/(4G_{_N})$.
In what follows, all the quantities are regarded as dimensionless
being rescaled by the parameter $\ell$.

Let us consider the boundary condition of the gauge fields.
What we would like to adopt is the boundary condition\cite{CHD,BBO}
\be\label{bc}
A_t = A_\phi ,\quad B_t = -B_\phi ,
\ee
at asymptotic infinity $\rho\sim\infty$.
This is the only appropriate boundary condition as will be seen later.
The variation of the Chern-Simons action is
\be\label{var}
\delta I_{CS}(B) - \delta I_{CS}(A) = \frac{k}{2\pi}{\rm tr}\int
\left(\delta BF_B - \delta AF_A\right)
+ \delta{\cal B}' ,
\ee
where
\be
\delta{\cal B}' = \frac{k}{4\pi}{\rm tr}\oint\left(
  \delta B_t B_\phi - \delta B_\phi B_t
- \delta A_t A_\phi + \delta A_\phi A_t
\right) .
\ee
What is needed is the boundary term ${\cal B}$ to be added to the
action $I_{CS}(B) - I_{CS}(A)$ such that the condition
$\delta{\cal B}+\delta{\cal B}'=0$ provides us with the boundary
condition (\ref{bc}).
The desirable boundary term turns out to be\cite{BBG}
\be\label{bt}
{\cal B} = \frac{k}{8\pi}{\rm tr}\oint
           \left(A_t^2 - A_\phi^2 + B_t^2 - B_\phi^2\right) .
\ee
We notice that the functional derivatives
$\delta/\delta A$ and $\delta/\delta B$ become well-defined in this
case.
In accordance with the boundary condition (\ref{bc}), the gauge
functions $u\in sl(2;{\bf R})_L$ and $v\in sl(2;{\bf R})_R$ also have
the boundary condition
$\partial_tu=\partial_\phi u$ and $\partial_tv=-\partial_\phi v$.
Adding the boundary term ${\cal B}$, the desirable action is
\be\label{L3}
L = I_{CS}(B) - I_{CS}(A) + {\cal B} .
\ee
Now, we postulate that the action of the system be $L$.
We will justify the boundary term in sect.~\ref{WZW}.

Let us find a solution of the action (\ref{L3}).
The more general solution will be found in sect.~\ref{WZW}.
The Einstein equation and torsion free condition are equivalently
expressed by the equations of motion
$F_A \equiv dA+A^2 = 0$ and $F_B \equiv dB+B^2 = 0$.
In this section we would like to find the BTZ black hole solution.
We will first find a constant solution $dA=dB=0$, and show
that this is gauge-transformed to the BTZ solution.
We tentatively use the gauge fixing condition $A_\rho=B_\rho=0$ only
for the purpose of finding a constant solution.
Now, the equations of motion are $[A_t,A_\phi]=[B_t,B_\phi]=0$.
These can be solved by setting $A_\phi = A_t$ and $B_\phi = -B_t$.
These relations are required from the boundary condition (\ref{bc}).
The quantities $A_t$ and $B_t$ may be any elements of the Lie algebra
$sl(2;{\bf R})$.
Each choice of the elements singles out various (black hole)
solutions.
For example, the case $A_t=-B_t=J_0$ corresponds to the anti-de-Sitter 
space $ds^2 = -(r^2+1)dt^2 + (r^2+1)^{-1}dr^2 + r^2d\phi^2$ as will be 
seen in sect.~\ref{WZW}.
For deriving the BTZ solution,
we take $A_t = z_+J_1$ and $B_t = z_-J_1$ by introducing
two arbitrary constants $z_\pm$.
Then, we obtain
\bea\label{degenerate}
A &=& \frac{z_+}{2}
        \pmatrix{0 & dt + d\phi \cr dt + d\phi & 0 \cr} ,\nonumber\\
B &=& \frac{z_-}{2}
        \pmatrix{0 & dt - d\phi \cr dt - d\phi & 0 \cr} .
\eea
Next, we perform the gauge transformation $g=\exp{\rho J_2}$ as
$A\to g^{-1}Ag + g^{-1}dg$ and $B\to gBg^{-1} + gdg^{-1}$, which is
belonging to $G/H$.
We have
\bea\label{solAB}
A &=& \frac{1}{2}
        \pmatrix{d\rho & z_+e^{-\rho}(dt + d\phi) \cr
                z_+e^\rho(dt + d\phi) & -d\rho \cr} \nonumber\\
  &=&  z_+(J_0\sinh\rho + J_1\cosh\rho)(dt+d\phi) + J_2d\rho
   ,\nonumber\\
B &=& \frac{1}{2}
        \pmatrix{-d\rho & z_-e^\rho(dt - d\phi) \cr
                z_-e^{-\rho}(dt - d\phi) & d\rho \cr} \nonumber\\
  &=&  z_-(J_0\sinh\rho - J_1\cosh\rho)(dt-d\phi) - J_2d\rho .
\eea
The tentative gauge fixing condition changes to
$A_\rho=J_2$ and $B_\rho=-J_2$.
It is interesting that radial functional form is introduced as
the broken gauge degrees of freedom.
Here we make a comment on $G/H$.
Unless we perform the gauge transformation $g\in G/H$,
the solution (\ref{degenerate}) would produce an unphysical degenerate
metric.
On the other hand, the metric determined by the solution (\ref{solAB}) 
is not degenerate, as is seen below, and has physical significance.
Namely, the gauge transformation $G/H$ connecting between the two
solutions has a physical meaning.
Expanding the above in terms of the generators (\ref{J_a}) and
using the relations $A = \omega +e$ and $B=\omega-e$,
we finally obtain
\be\label{BTZsol}
\begin{array}{ll}
e^0 = \sinh\rho(r_+dt- r_-d\phi) ,&
   \omega^0 = \sinh\rho(r_+d\phi- r_-dt) ,\\
e^1 = \cosh\rho(r_+d\phi- r_-dt) ,&
    \omega^1 = \cosh\rho(r_+dt- r_-d\phi),\\
e^2 = d\rho ,&
    \omega^2 = 0 ,
\end{array}
\ee
where we put $z_\pm = \pm r_+ - r_-$.
Reparameterizing the radial coordinate as
$r^2 = r_+^2\cosh^2\rho-r_-^2\sinh^2\rho$ allows us to obtain the
standard form of the BTZ black hole geometry\cite{BTZ}
\bea\label{BTZmetric}
ds^2 &=& \eta_{ab}e^ae^b \nonumber\\
  &=& -N^2dt^2 + N^{-2}dr^2 + r^2(d\phi+N_\phi dt)^2 ,\nonumber\\
&& N^2 = \frac{(r^2-r^2_+)(r^2-r^2_-)}{r^2} ,\nonumber\\
&& N_\phi = -\frac{r_+r_-}{r^2} .
\eea
If we regard $r_+$ and $r_-$ as the outer and inner horizons,
respectively, then we have $r_+\ge r_->0$.
We notice that $\rho=0$ corresponds to the event horizon $r=r_+$.

The solution can be extended to inside of the event
horizon.\cite{BHTZ}
However, since we are interested in the asymptotic behavior of the
system at infinity, we will concentrate on the outside of the event
horizon.

\section{Asymptotic symmetry at infinity}
\label{ASAI}
\reseteqnum

In the Chern-Simons theory the diffeomorphism of the gauge field is
embedded into the gauge transformation.
The embedded diffeomorphism becomes an on-shell symmetry.
Since the background satisfies the on-shell condition, the embedded
diffeomorphism makes sense when we consider the Brown-Henneaux
asymptotic symmetry.
In the present system, under consideration, the diffeomorphisms of the 
gauge fields $A$ and $B$ are embedded into the gauge group
$SL(2;{\bf R})_L$ and $SL(2;{\bf R})_R$, respectively.
What we will show is that the Brown-Henneaux asymptotic symmetry is
not merely this usual diffeomorphism but a superposition of the usual
diffeomorphism and a gauge transformation.
We would like to emphasize that the isometry is given by
\be\label{embeddiffeo}
u^a = \xi^mA^a_m + w^a ,\quad v^a = \xi^mB^a_m + w^a ,
\ee
where $w^a$ is a gauge function depending on $\xi^m$.
This means that the isometry is not the usual diffeomorphism in the
Chern-Simons formulation.
The gauge functions (\ref{embeddiffeo}) are in a sharp contrast with
the prescription used in refs.~\cite{Banados,BBO}.
Unless we consider the gauge function $w^a$ in (\ref{embeddiffeo}),
one would be forced to introduce two kinds of diffeomorphism parameter 
$\xi^m$ in $u^a$ and $v^a$ separately.
This is hardly acceptable.
It should be stressed that our result (\ref{embeddiffeo}) is the only
consistent way to explain the Brown-Henneaux asymptotic symmetry in
the Chern-Simons formulation.
In the following we derive the form (\ref{embeddiffeo}) in more
detail.

The diffeomorphism, with parameter $\xi^m$, of one-form is given by
$\delta_{_D} A_m = \xi^n\partial_nA_m+\partial_m\xi^nA_n$.
At on-shell $F_{mn}=0$ this can be embedded into the gauge
transformation.
The infinitesimal gauge transformations of the left- and right-handed
sectors are given by
$\delta_{_L} A=du+[A,u]$ and
$\delta_{_R} B=dv+[B,v]$,
so the gauge transformations for the dreibein and spin connection are
\bea\label{gaugeew}
\delta_{_G} e^a_m &=& \frac{1}{2}\left(
    \delta_{_L}A^a_m-\delta_{_R}B^a_m \right)
  = \partial_m u_-^a
  + \epsilon^a{}_{bc}(\omega_m^bu_-^c + e^b_m u_+^c ) ,\nonumber\\
\delta_{_G} \omega^a_m &=& \frac{1}{2}\left(
    \delta_{_L}A^a_m+\delta_{_R}B^a_m \right)
  = \partial_m u_+^a
  + \epsilon^a{}_{bc}(\omega_m^bu_+^c + e^b_m u_-^c ) ,
\eea
where $u^a_\pm = (u^a \pm v^a)/2$.
Note that we are using the quantities $u^a$ and $v^a$ as
generic $SL(2;{\bf R})_L$ and $SL(2;{\bf R})_R$ gauge functions,
respectively.
The on-shell diffeomorphisms of the gauge fields are expressed by
$u^a = \xi^mA^a_m$ and $v^a = \xi^mB^a_m$, respectively.
These gauge functions $u^a$ and $v^a$ define the embedded
diffeomorphism of the dreibein $\delta_{_D}e^a_m$ and the spin
connection $\delta_{_D}\omega^a_m$ through eqs.~(\ref{gaugeew}).
The diffeomorphism of the dreibein $\delta_{_D}e^a_m$ induces that of
the metric $g_{mn}=\eta_{ab}e^a_me^b_n$;
\be
\delta_{_D}g_{mn} =
\xi^l\partial_lg_{mn}+\partial_m\xi^lg_{ln}+\partial_n\xi^lg_{ml} .
\ee
So, one might think that the isometry group defined by
$\delta_{_D}g_{mn}=0$ would be found by the condition
$\delta_{_D}e^a_m=0$.
However, this is not correct.
We have to take account of the gauge degrees of freedom in $H$.

The gauge transformation of the dreibein (\ref{gaugeew}) induces that
of the metric
\be
\delta_{_G}g_{mn} = \eta_{ab}(e^a_mD_nu_-^b+e^a_nD_mu_-^b) ,
\ee
where $D_mu_-^a \equiv \partial_m u_-^a +
\epsilon^a{}_{bc}\omega^b_mu_-^c$.
Since the metric is not a gauge invariant quantity of the gauge
group $G$, the presence of the metric triggers the spontaneous
symmetry breaking of the gauge symmetry $G$.
The presence of the metric, namely in a broken phase, does not mean
the breaking of the whole gauge symmetry $G$, but preserves a part of
the symmetry $H\subset G$.
The invariance of the metric $\delta_{_G}g_{mn}=0$ means
\be\label{u=v}
u=v ,
\ee
which defines the manifest symmetry $H$ in the broken phase.
Namely, the metric is still invariant under the gauge transformation
\be\label{manifestsym}
\delta_{_H}e^a_m = \epsilon^a{}_{bc}e^b_mw^c ,
\ee
where we put $u_- = 0$ and $u_+ = w$ in the transformation
(\ref{gaugeew}).
This is an $SO(2,1)$ rotation, with parameter $w$, in the local
frame.
The manifest symmetry $H$ is recognized as the vector-like subgroup
$SL(2;{\bf R})_{L+R}$ defined by eq.~(\ref{u=v}).

In conclusion, the condition of the isometry $\delta_{_D}g_{mn}=0$
should be interpreted, in the dreibein formulation, as the condition
\be\label{isometry}
\delta_{_D}e^a_m + \delta_{_H}e^a_m
  = \xi^n\partial_ne^a_m+\partial_m\xi^ne^a_n
  + \epsilon^a{}_{bc}e^b_mw^c
  = 0 .
\ee
Of course, if we did not use the dreibein, the condition
$\delta_{_D}g_{mn}=0$ determines $\xi^m$ consistently.
However, the gauge degrees of freedom $w^a$ could not be determined.
The condition (\ref{isometry}) determines both of the diffeomorphism
function $\xi^m$ and associating gauge function $w^a$ in
eqs.~(\ref{embeddiffeo}).

Next, let us solve the isometry condition (\ref{isometry}).
By using the concrete form of the dreibein (\ref{BTZsol}),
the isometry condition becomes
\bea\label{isocon}
&\displaystyle \frac{\hat{g}_{jk}}{(r_+^2-r_-^2)\cosh\rho\sinh\rho}
   \,\partial_i\xi^k + \xi^\rho\eta_{ij}
           - w^2\epsilon_{ij} = 0 ,&\nonumber\\
& w^a = - \epsilon^a{}_be^b_i\partial_\rho\xi^i ,&\nonumber\\
& \partial_i\xi^\rho = -\hat{g}_{ij}\partial_\rho\xi^j ,&
\eea
where $a,b=0,1$, $\epsilon^{ab}\equiv\epsilon^{ab2}$, $i,j=t,\phi$,
$\xi_i \equiv \eta_{ij}\xi^j$ and
\be
\hat{g}_{ij} =
\pmatrix{r_-^2\cosh^2\rho-r_+^2\sinh^2\rho & -r_+r_- \cr
         -r_+r_- & r_+^2\cosh^2\rho-r_-^2\sinh^2\rho \cr} .
\ee
Let us investigate the isometry condition (\ref{isocon}) at infinity
$\rho\sim\infty$.
In the leading order in $e^\rho$ it becomes 
\bea\label{isoinfty}
& \partial_i\xi_j + \xi^\rho\eta_{ij}
           - w^2\epsilon_{ij} = 0 ,&\nonumber\\
&\displaystyle \partial_\rho(\xi^t\pm\xi^\phi) =
           \frac{8e^{-2\rho}}{r_+^2-r_-^2}\partial_\mp\xi^\rho
            ,&\nonumber\\
&\displaystyle w^0\pm w^1 = \frac{\pm r_+ - r_-}{2}e^\rho
           \partial_\rho(\xi^t\pm\xi^\phi) .&
\eea
Especially, the symmetric part of the first condition is
\be
\partial_i\xi_j + \partial_j\xi_i + 2\xi^\rho\eta_{ij} = 0 .
\ee
This is the condition for the conformal Killing vector in the
asymptotic region at infinity.
We should note that if we did not include the gauge symmetry
contribution (\ref{manifestsym}), say $w=0$, we would not obtain
the conformal symmetry.
The solution of the condition (\ref{isoinfty}) is
\bea\label{isosol}
\xi^t\pm\xi^\phi &=& 2T_\pm +
\frac{2e^{-2\rho}}{r_+^2-r_-^2}\partial_\mp^2T_\mp,\nonumber\\
\xi^\rho &=& -(\partial_+T_+ + \partial_-T_-) ,\nonumber\\
w^0\pm w^1 &=& - \frac{4e^{-\rho}}{\pm r_+ + r_-}
                 \partial_\mp^2T_\mp ,\nonumber\\
w^2 &=& -\partial_+T_+ + \partial_-T_- ,
\eea
where $T_\pm = T_\pm(t\pm\phi)$ are arbitrary function.

We may also hope to investigate the isometry condition
(\ref{isocon}) near horizon $\rho\sim0$.
However, the first condition of eqs.~(\ref{isocon})
diverges, and we should change the coordinates to regular ones like
Kruskal coordinates.
The condition (\ref{isocon}) is not so easily solved in general
$\rho$, but on any slice surface with constant $\rho$ the isometry is
determined by the condition (\ref{isocon}).
As will be shown in sect.~\ref{WZW}, this is manifestly seen by the
fact that the boundary theory of the Chern-Simons theory is described
by the Wess-Zumino-Witten action.

\section{Current algebras of Chern-Simons theory}
\label{CACS}
\reseteqnum

Following refs.~\cite{BH,RT,Banados} we define the first class
constraint in the presence of the boundary.
In this section we derive the Kac-Moody currents and the Virasoro
generators making good use of the asymptotic symmetry
(\ref{embeddiffeo}).

The Chern-Simons action (\ref{L3}) defines the Poisson's brackets
\bea
\{A^a_i(x),A^b_j(y)\}_{_P} &=&
-\frac{4\pi}{k}\epsilon_{ij}\eta^{ab}\delta^2(x-y) ,\nonumber\\
\{B^a_i(x),B^b_j(y)\}_{_P} &=&
\frac{4\pi}{k}\epsilon_{ij}\eta^{ab}\delta^2(x-y) ,\nonumber\\
\{A^a_i(x),B^b_j(y)\}_{_P} &=& 0 ,
\eea
where $i,j = \phi, \rho$ and $x,y \in {\bf R}\times \partial\Sigma$.
The first class constraint is $F_{Aij}\approx F_{Bij}\approx 0$,
and their integrated forms are
\be
g_{_A}(\Lambda) = \frac{k}{4\pi}{\rm tr}\int_\Sigma
\epsilon^{ij}F_{Aij}\Lambda
\ee
and similarly for $g_{_B}(\Lambda)$.
For later convenience the gauge function $\Lambda$ does or does not
depend on the gauge fields.
In order to calculate the Poisson's bracket of the first class
constraint, they should be varied properly even on the boundary.
Then, it is necessary to add a boundary term $Q_{_A}(\Lambda)$ as
$g_{_A}(\Lambda) \to G_{_A}(\Lambda) \equiv g_{_A}(\Lambda) +
Q_{_A}(\Lambda)$ and similarly for $g_{_B}(\Lambda)$.\cite{Banados}
The boundary terms are defined, in the variational form, by
\bea\label{dQ}
\delta Q_{_A}(\Lambda) &=& - \frac{k}{2\pi}{\rm tr}\int_\Sigma
\epsilon^{ij}\partial_i(\delta A_j\Lambda) ,\nonumber\\
\delta Q_{_B}(\Lambda) &=& - \frac{k}{2\pi}{\rm tr}\int_\Sigma
\epsilon^{ij}\partial_i(\delta B_j\Lambda) .
\eea
Then, weakly we have
\bea
\{A^a_i(x),G_{_A}(\Lambda)\}_{_P} &\approx&
  \partial_i\Lambda^a + \epsilon^a{}_{bc}A^b_i\Lambda^c ,\nonumber\\
\{B^a_i(x),G_{_B}(\Lambda)\}_{_P} &\approx&
  -\partial_i\Lambda^a - \epsilon^a{}_{bc}B^b_i\Lambda^c .
\eea
Since the boundary term $Q_{_A}$ is weakly equivalent to $G_{_A}$,
we may define the Poisson's bracket between the boundary terms as
\be
\{Q_{_A}(\alpha),Q_{_A}(\beta)\}_{_P} \equiv
\left.\{G_{_A}(\alpha),G_{_A}(\beta)\}_{_P}\right|_{g_A\approx0}
\ee
and similarly for $Q_{_B}$.

First, let us derive the generators of the gauge transformation and
their Poisson's bracket relations; namely the Kac-Moody algebra.
If the gauge function $\Lambda$ does not depend on the gauge field,
this is immediately integrated to be
\bea
Q_{_A}(\Lambda)
&=& - \frac{k}{2\pi}{\rm tr}\int_\Sigma
  \epsilon^{ij}\partial_i(A_j\Lambda) 
= \frac{k}{2\pi}\int_0^{2\pi}\!\!d\phi\:
  {\rm tr}(A_\phi\Lambda)  ,\nonumber\\
Q_{_B}(\Lambda)
&=& \frac{k}{2\pi}\int_0^{2\pi}\!\!d\phi\:
  {\rm tr}(B_\phi\Lambda)  ,
\eea
where we suppressed the integration constant.
Notice that $\epsilon^{\rho\phi}=-1$.
Therefore, we obtain
\bea
\{Q_{_A}(\alpha),Q_{_A}(\beta)\}_{_P} &=&
- Q_{_A}([\alpha,\beta])
- \frac{k}{2\pi}{\rm tr}\int_\Sigma
\epsilon^{ij}\partial_i(\alpha\partial_j\beta) ,\nonumber\\
\{Q_{_B}(\alpha),Q_{_B}(\beta)\}_{_P} &=&
Q_{_B}([\alpha,\beta])
+ \frac{k}{2\pi}{\rm tr}\int_\Sigma
\epsilon^{ij}\partial_i(\alpha\partial_j\beta) .
\eea
Let us expand the gauge fields in terms of the Fourier modes $J^a_n$
and $\overline{J}^a_n$ as
\bea\label{Fourier}
A^a_\phi &=& -\frac{2}{k}\sum_n J_n^ae^{-in(t+\phi)}
  ,\nonumber\\
B^a_\phi &=& \frac{2}{k}\sum_n \overline{J}_n^ae^{-in(t-\phi)} .
\eea
Here, the overall coefficients in eqs.~(\ref{Fourier}) are determined
such that the resulting Kac-Moody algebras take the  standard form.
The reality condition of the gauge fields
$(A^a_m)^\dagger = A^a_m$ and $(B^a_m)^\dagger = B^a_m$
implies
$(J^a_n)^\dagger =J^a_{-n}$ and
$(\overline{J}^a_n)^\dagger =\overline{J}^a_{-n}$.
The Kac-Moody algebra reads
\bea\label{KMalg}
i\{J^a_n,J^b_m\}_{_P} &=& i\epsilon^{ab}{}_c J^c_{n+m}
  + \frac{k}{2}n\eta^{ab}\delta_{n+m} ,\nonumber\\
i\{\overline{J}^a_n,\overline{J}^b_m\}_{_P}
  &=& i\epsilon^{ab}{}_c \overline{J}^c_{n+m}
  + \frac{k}{2}n\eta^{ab}\delta_{n+m} .
\eea
Here we make a comment on the definition of the currents
$J^a_n$ and $\overline{J}^a_n$.
The $\rho$ dependence of the gauge field components (\ref{Fourier})
arises through the  homogeneous gauge transformation belonging to
$G/H$.
Strictly speaking, this dependence should be eliminated from the
currents by performing an inverse homogeneous gauge transformation.
For details, the reader is referred to eqs.~(\ref{puregauge}) and
(\ref{gpm}).

Next, let us derive the generators of the isometry embedded into the
gauge symmetry and their Poisson's bracket relations;
namely the Virasoro algebra.
The gauge fields satisfy the boundary condtion (\ref{bc}) and
the gauge fixing condition $A_\rho = J_2$ and $B_\rho = -J_2$.
So, the twisted diffeomorphism (\ref{embeddiffeo}) with
(\ref{isosol}) becomes
\bea\label{as}
u &=&  2T_+A_\phi - 2\partial_+T_+\alpha ,\nonumber\\
v &=& -2T_-B_\phi + 2\partial_-T_-\alpha ,
\eea
where we define $\alpha=J_2$.
It is a subtle fact that these quantities $u$ and $v$ satisfy the
boundary condition, although the components $\xi^\rho$ and $w^2$ in
the solution (\ref{isosol}) do not.
Then, after the integration of eq.~(\ref{dQ}), we obtain
\bea
Q_{_A}(\xi) &=&   \frac{k}{2\pi}\int_0^{2\pi}\!\!d\phi\:
  T_+{\rm tr}\left(A_\phi^2+2\alpha\partial_+A_\phi\right)
  ,\nonumber\\
Q_{_B}(\xi) &=& - \frac{k}{2\pi}\int_0^{2\pi}\!\!d\phi\:
  T_-{\rm tr}\left(B_\phi^2+2\alpha\partial_-B_\phi\right) .
\eea
On putting $T_\pm = e^{in(t\pm\phi)}$, these boundary terms define
the Virasoro generators
\bea\label{VirasoroGen}
L_n &=& \frac{1}{2}Q_{_A}(\xi)
    = \frac{1}{k}\sum_mJ_{am}J^a_{n-m} + in\alpha_aJ^a_n + a_0\delta_n
  ,\nonumber\\
\overline{L}_n &=& -\frac{1}{2}Q_{_B}(\xi)
    = \frac{1}{k}\sum_m\overline{J}_{am}\overline{J}^a_{n-m}
    - in\alpha_a\overline{J}^a_n + a_0\delta_n ,
\eea
where we added a constant $a_0=k/4$.
We see that the Virasoro generators are not the usual energy-momentum
tensor, i.e., not the Sugawara form, but the twisted energy-momentum
tensor
\bea\label{twist}
T &\to& T + \partial_+ J^2 ,\nonumber\\
\overline{T} &\to& \overline{T}
     + \partial_-\overline{J}^2 .
\eea
Using eqs.~(\ref{KMalg}), we arrive at the Poisson's bracket relations 
\bea
i\{L_n,L_m\}_{_P} &=& (n-m)L_{n+m} +\frac{k}{2}n(n^2-1)\delta_{n+m}
  ,\nonumber\\
i\{\overline{L}_n,\overline{L}_m\}_{_P}
  &=&
  (n-m)\overline{L}_{n+m} +\frac{k}{2}n(n^2-1)\delta_{n+m} .
\eea
Now, we have reproduced the Brown-Henneaux asymptotic Virasoro algebra
with the central charge $6k$.

\section{General solution of the Chern-Simons theory and Relation to
 the Wess-Zumino-Witten model}
\label{WZW}
\reseteqnum

In this section we derive the general solution of the Chern-Simons
theory (\ref{L3}) and show its relation to the boundary theory.
Let us solve the equations of motion $F_A=F_B=0$ under the
appropriate boundary condition (\ref{bc}).
Thanks to the field equations, the gauge fields become pure gauge form 
\be\label{puregauge}
A_m=g^{-1}\partial_mg ,\quad
B_m=-\partial_m\overline{g}\,\overline{g}^{-1} .
\ee
Constrained by the boundary condition (\ref{bc}),
the group elements $g(t,\phi,\rho)$ and $\overline{g}(t,\phi,\rho)$
become functions of $t+\phi$ and $t-\phi$ only on the boundary,
respectively; $h(t+\phi)$ and $\overline{h}(t-\phi)$.
Such group elements define the space of connection ${\cal A}$.
Next, we would like to divide the space of connections ${\cal A}$ by
the space of gauge functions.
Let us define the space of gauge function $\hat{G}_0$ such that any
element in $\hat{G}_0$ becomes the identity on the boundary.
Now, the space of gauge inequivalent classes is ${\cal A}/\hat{G}_0$.
If we restrict these spaces ${\cal A}$ and $\hat{G}_0$ inside the
boundary, these spaces become exactly the same.
Then, the quotient space ${\cal A}/\hat{G}_0$ will be expected as the
space of the boundary degrees of freedom
$h(t+\phi)$ and $\overline{h}(t-\phi)$.
Let us show this fact.
The gauge degrees of freedom $\hat{G}_0$ are equally fixed by imposing 
the gauge fixing condition
\be\label{gf}
A_\rho = \beta+\alpha ,\quad B_\rho = \beta-\alpha ,
\ee
where $\alpha$ and $\beta$ are any constant element of
$SL(2;{\bf R})$.
This follows from the fact that the residual gauge transformation
\bea
A_\rho &=& f^{-1}A_\rho f + f^{-1}\partial_\rho f ,\nonumber\\
B_\rho &=& \overline{f}^{-1}B_\rho\overline{f}
       + \overline{f}^{-1}\partial_\rho\overline{f} ,
\eea
is easily integrated giving the elements of $\hat{G}_0$ as
$f=\overline{f}=1$.
After imposing the gauge fixing condition, the group elements $g$ and
$\overline{g}$ take a form $g=h(t,\phi)e^{(\alpha+\beta)\rho}$ and
$\overline{g}=e^{(\alpha-\beta)\rho}\overline{h}(t,\phi)$.
Now, imposing the boundary condition (\ref{bc}),
we finally obtain
\be\label{gpm}
g=h(t+\phi)e^{(\alpha+\beta)\rho} ,\quad
\overline{g}=e^{(\alpha-\beta)\rho}\overline{h}(t-\phi) .
\ee
At first sight one might think that since the solution
(\ref{puregauge}) took a pure gauge form, the solution would be gauged 
away completely.
However, our result (\ref{gpm}) illustrates that the boundary degrees
of freedom $h(t+\phi)$ and $\overline{h}(t-\phi)$ survive as the
physical degrees of freedom of the system.
These group elements with eqs.~(\ref{puregauge}) give quite general
solutions of the Chern-Simons gravity.
Since the metric should not degenerate, $\alpha\ne0$ is assumed.
In this situation, the $g_{\rho\rho}$ part of the metric can be set
equal to unity $g_{\rho\rho}=1$ by rescaling the radial variable
$\rho$, and then we may put $\alpha_a\alpha^a=1$.
For later simplicity we set $\alpha=J_2$ and $\beta=0$.

According to the solution (\ref{gpm}), the dreibein and
spin-connection become
\be
\begin{array}{ll}
\displaystyle
e_+ = \frac{1}{2}e^{-\alpha\rho}h^{-1}\partial_+he^{\alpha\rho} ,&
\displaystyle
\medskip\omega_+ = e_+ ,\\
\displaystyle
e_- = \frac{1}{2}e^{\alpha\rho}\partial_-
      \overline{h}\,\overline{h}^{-1}e^{-\alpha\rho} ,&
\displaystyle
\omega_- = -e_- ,\\
\displaystyle
e_\rho = \alpha ,&
\displaystyle
\omega_\rho = 0 ,
\end{array}
\ee
where the subindices $\pm$ indicate that quantities $f_\pm$ are the $dt\pm
d\phi$ components of one-forms $f=e^a, \omega^a, d$.
This is the most general gauge-fixed solution of the system
(\ref{L3}).
The black hole metric becomes
\bea\label{generalsol}
g_{++} &=& \frac{1}{2}{\rm tr}\left(h^{-1}\partial_+h\right)^2
           ,\nonumber\\ 
g_{--} &=& \frac{1}{2}{\rm tr}\left(
  \partial_-\overline{h}\,\overline{h}^{-1}\right)^2,\nonumber\\ 
g_{+-} &=& \frac{1}{2}{\rm tr}\left(
  e^{-2\alpha\rho}h^{-1}\partial_+h
  e^{ 2\alpha\rho}\partial_-\overline{h}\,\overline{h}^{-1}
  \right),\nonumber\\
g_{\rho+} &=& {\rm tr}\left(
  \alpha h^{-1}\partial_+h\right),\nonumber\\ 
g_{\rho-} &=& {\rm tr}\left(
  \alpha\partial_-\overline{h}\,\overline{h}^{-1}
  \right),\nonumber\\ 
g_{\rho\rho} &=& 1 ,
\eea
where the trace is taken on the subgroup $H$.
At infinity $\rho\sim\infty$,
the asymptotic behavior of the metric is
\be
g_{+-} \sim {\cal O}(e^{2\rho})
\ee
and the others are of ${\cal O}(1)$.
This behavior provides us with the same isometry solution
(\ref{isosol}) again.
The solution (\ref{generalsol}) shows that the metric is specified by
an infinite number of conserved currents, i.e., the Kac-Moody
currents.
We note that the BTZ black hole\cite{BTZ} corresponds to the special
choice
\be\label{BTZsol2}
h = e^{z_+J_1\cdot(t+\phi)} ,\quad
\overline{h} = e^{z_-J_1\cdot(t-\phi)} .
\ee
And the vacuum solution corresponding to the anti-de-Sitter space
is not represented by (\ref{BTZsol2}) for any $z_\pm\in{\bf R}$ but
is given by
\be
h = e^{J_0\cdot(t+\phi)} ,\quad
\overline{h} = e^{J_0\cdot(t-\phi)} ,
\ee
by which we find
\bea
e &=& J_0\cosh\rho dt + J_1\sinh\rho d\phi + J_2d\rho ,\nonumber\\
w &=& J_0\cosh\rho d\phi + J_1\sinh\rho dt .
\eea

The two dimensional parts $h(t+\phi)$ and $\overline{h}(t-\phi)$ of
the solution (\ref{gpm}) also can be given by the Wess-Zumino-Witten
action
\be
I_\pm(h) =
-\frac{k}{4\pi}{\rm tr}\oint\partial_+h^{-1}\partial_-h
\mp\frac{k}{12\pi}{\rm tr}\int\left(h^{-1}dh\right)^3 ,
\ee
and, then, the action of the boundary theory would become
$L = I_+(h) + I_-(\overline{h})$.
Unfortunately, the symmetry of this action is twice as large as that
of the boundary theory.
We have to throw away the modes represented by $-\partial_-hh^{-1}$
and $\overline{h}^{-1}\partial_+\overline{h}$.
Instead of this choice, we have more excellent one.
Suppose that the representations of the left- and right-handed
generators are the same.
Let us consider the Wess-Zumino-Witten theory $I_+(g)$ with the
subgroup gauge symmetry $H$.
Then, the field equation
$\partial_-(g^{-1}\partial_+g)=0$
provides us with the two conserved currents 
$g^{-1}\partial_+g=h^{-1}\partial_+h$ and
$-\partial_-gg^{-1}=-\partial_-\overline{h}\,\overline{h}^{-1}$, 
by regarding the identification $g=\overline{h}(t-\phi)h(t+\phi)$.
Then, in this situation the action of the boundary theory becomes
\be\label{L2}
L = I_+(g) .
\ee

Finally, let us show that after imposing the constraints the boundary
term in $\delta L$, namely
\be
0 = \frac{k}{4\pi}{\rm tr}\oint\left(
\delta A_+A_- + \delta B_-B_+ \right) ,
\ee
becomes the field equation of the boundary theory (\ref{L2}).

The constraint is given by the variation of $L$ with respect to the
multipliers $A_t$ and $B_t$, so we have
$F_{_A}{}_{\rho\phi}=F_{_B}{}_{\rho\phi}=0$.
These are easily integrated to be
$A_\phi=e^{-\alpha\rho}J(t,\phi)e^{\alpha\rho}$ and
$B_\phi=e^{\alpha\rho}\overline{J}(t,\phi)e^{-\alpha\rho}$.
Since the quantity $J(t,\phi)$ ($\overline{J}(t,\phi)$) should be
a $\phi$ component of a one-form and be Lie algebra valued,
they should be given by the Maurer-Cartan one-forms
$J = h^{-1}\partial_\phi h$ and
$\overline{J} = -\partial_\phi\overline{h}\,\overline{h}^{-1}$
for some appropriate group elements $h(t,\phi)$ and
$\overline{h}(t,\phi)$.
Since $A_\rho$ and $B_\rho$ are gauge-fixed, we have to impose the
Gauss law constraints
$\delta L/\delta A_\rho=\delta L/\delta B_\rho=0$ which lead to the
conditions
\bea
\partial_tJ &=& \partial_\phi K + [J,K] ,\nonumber\\
\partial_t\overline{J} &=&
  \partial_\phi \overline{K} +
  [\overline{J},\overline{K}] ,
\eea
where we put
$K = e^{-\alpha\rho}A_te^{\alpha\rho}$ and
$\overline{K} = e^{\alpha\rho}B_te^{-\alpha\rho}$.
These determine the multipliers $A_t$ and $B_t$ as
$K = h^{-1}\partial_t h$ and
$\overline{K} = -\partial_t\overline{h}\,\overline{h}^{-1}$.
We note that the variations with respect to $\delta A_\phi$ and
$\delta B_\phi$ are the field equation, which becomes identity
equation after imposing the constraints.
Now, using the above constraints, the boundary condition from the
boundary term is equivalent to
\be
0 = \frac{k}{4\pi}{\rm tr}\oint\left[
h^{-1}\delta h\partial_-\left(h^{-1}\partial_+h\right)
 +
   \delta\overline{h}\,\overline{h}^{-1}\partial_+
   \left(\partial_-\overline{h}\,\overline{h}^{-1}\right)
\right] .
\ee
This is nothing but the variational equation of the Wess-Zumino-Witten
action (\ref{L2}); $\delta I_+(g) = 0$ with the identification
of the solution $g(t,\phi) = \overline{h}(t-\phi)h(t+\phi)$.
In other words, after imposing the constraints, the boundary condition 
of the bulk theory becomes the field equation of the boundary theory.

In conclusion, the boundary CFT theory of the $AdS_3$ gravity theory
(\ref{L3}) is equivalent to the Wess-Zumino-Witten theory (\ref{L2}).
The usual energy-momentum tensor takes the Sugawara form as is easily
derived by the Noether method.
However, this is not exactly the generator of the asymptotic symmetry, 
but merely a part of it.
The correct generator is the energy-momentum tensor
(\ref{VirasoroGen}) twisted by the contribution of the local rotation
by the vector-like gauge subgroup $SL(2;{\bf R})_{L+R}$.

\section{Quantization of Chern-Simons theory}
\label{QCS}
\reseteqnum

When the system is quantized, the Poisson's bracket is replaced by the
commutator bracket; $i\{~,~\}_{_P}\to[~,~]$.
Immediately, we obtain the two Kac-Moody algebras of center $k$.
In quantized version, the coefficient of the linear term is modified
in the twisted energy-momentum tensor
\bea
T &\to& T + \frac{k}{k-2}\partial_+ J^2
    ,\nonumber\\
\overline{T} &\to& \overline{T}
    + \frac{k}{k-2}\partial_-\overline{J}^2 .
\eea
As usual, the currents in the definition of the Virasoro generators
(\ref{VirasoroGen}) should be normal ordered like
\be
:J^a_nJ^b_m: \ = \cases{
               J^a_nJ^b_m & ($n \le m$) \cr
               J^b_mJ^a_n & ($m < n$) \cr
               } ,
\ee
and similarly for $\overline{J}^a_n$.
Then, the definition of the Virasoro generators are modified to be
\bea\label{VirasoroGen2}
L_n &=& \frac{1}{k-2}\sum_m:J_{am}J^a_{n-m}:
      + \:\frac{k}{k-2}in\alpha_aJ^a_n + a_0\delta_n
  ,\nonumber\\
\overline{L}_n &=&
      \frac{1}{k-2}\sum_m:\overline{J}_{am}\overline{J}^a_{n-m}:
      - \:\frac{k}{k-2}in\alpha_a\overline{J}^a_n + a_0\delta_n,
\eea
where we added a constant $a_0=\frac{k^3}{4(k-2)^2}$.
The commutation relations are
\bea
{}[L_n,L_m] &=& (n-m)L_{n+m} +\frac{c}{12}n(n^2-1)\delta_{n+m}
  ,\nonumber\\
{}[\overline{L}_n,\overline{L}_m] &=&
(n-m)\overline{L}_{n+m} +\frac{c}{12}n(n^2-1)\delta_{n+m} ,
\eea
where
\be\label{c}
c = \frac{3k}{k-2} + 6k\left(\frac{k}{k-2}\right)^2 .
\ee
At weak coupling limit $k\to\infty$ the center (\ref{c}) takes the
classical value $6k$ in the leading order.

The physical state condition is
\bea
& J^a_0\vert{\rm phys}\rangle = \vert{\rm phys}\rangle J^a ,
& L_0\vert{\rm phys}\rangle = h_0\vert{\rm phys}\rangle , \nonumber\\
& J^a_n\vert{\rm phys}\rangle = 0 ,~~~~~~~~~
& L_n\vert{\rm phys}\rangle = 0 ,
\eea
for $n = 1, 2, 3, \cdots$, where
$h_0=J_aJ^a/(k-2)+a_0$.

\section{Summary and discussions}\label{DIS}
\reseteqnum

We have analyzed the Chern-Simons gravity theory with the boundary
term (\ref{bt}).
We introduced the boundary term ${\cal B}$ in the action
$I_{CS}(B) - I_{CS}(A)$ such that the condition
$\delta{\cal B}+\delta{\cal B}'=0$ uniquely provides us with the
boundary condition (\ref{bc}).
If we did not introduce the boundary term,
the boundary theory would be a pair of the chiral Wess-Zumino-Witten
theory
\be
I_{c\pm}(g) =
-\frac{k}{4\pi}{\rm tr}\oint\partial_tg^{-1}\partial_\phi g
\mp\frac{k}{12\pi}{\rm tr}\int\left(g^{-1}dg\right)^3 ,
\ee
in which theory the Kac-Moody currents would not be guaranteed to be
a function only of $t\pm\phi$.
This theory does not describe the asymptotic behavior of the
Chern-Simons theory with the boundary condition (\ref{bc}).
In other words, without the boundary term ${\cal B}$,
the gauge functions $u$ and $v$ for the asymptotic symmetry (\ref{as})
would not be guaranteed to consist of the left- and right-moving
modes, respectively, since $A_\phi$ and$B_\phi$ is not a function of
$t+\phi$ and $t-\phi$, respectively.
This result mismatches the fact that the asymptotic symmetry (\ref{as})
is a function of $t+\phi$ or $t-\phi$.

The Brown-Henneaux asymptotic symmetry should be exclusively derived
from the condition (\ref{isometry}).
The most important point is that the diffeomorphism should be
accompanied by the manifest gauge transformation in $H$ through
eqs.~(\ref{embeddiffeo}) in the Chern-Simons formulation.
In our formulation the gauge functions $u$ and $v$ for the asymptotic
symmetry (\ref{as}) are naturally made of the left- and right-moving
modes, respectively.
Unless we consider the gauge function $w^a$ in (\ref{embeddiffeo}),
one would be forced to introduce two kinds of diffeomorphism parameter 
$\xi^m$ in $u^a$ and $v^a$ separately.\cite{Banados,BBO}
This is unfavorable.

We have found the general solution (\ref{gpm}) of the $AdS_3$
Chern-Simons gravity.
Dividing the space of solutions ${\cal A}$ by the space of gauge
transformations $\hat{G}_0$, the resultant general solution is
characterized by $h(t+\phi)$ and $\overline{h}(t-\phi)$ which are
boundary degrees of freedom of the bulk theory.

The physical boundary degrees of freedom $h(t+\phi)$ and
$\overline{h}(t-\phi)$ of the $AdS_3$ gravity theory (\ref{L3}) are
described by the Wess-Zumino-Witten theory (\ref{L2}).
Then, we have written down the quite general form of the metric
(\ref{generalsol}) in terms of the Kac-Moody currents.
The BTZ black hole is included in our black hole solutions.

The classical central charge of the Virasoro algebra is $6k$,
while the quantum counterpart becomes eq.~(\ref{c}).

\begin{flushleft}
\bf Acknowledgements
\end{flushleft}
The work of T.K. was supported in part by Scientific Grants from the
Ministry of Education (Grant Number, 09640353).


\end{document}